# Convection and intermittency noise in water temperature near a deep Mediterranean seafloor

**by Hans van Haren**

NIOZ Royal Netherlands Institute for Sea Research, P.O. Box 59, 1790 AB Den Burg, the Netherlands.
e-mail: hans.van.haren@nioz.nl

ABSTRACT

Turbulent and internal wave motions are important for the exchange of momentum, heat and suspended matter in the deep-sea which is generally stably stratified in density. Turbulence-generation models involve shear of vertical current differences that deforms the stratified waters, and convection that is drive by (unstable) buoyancy. Shear-generation is found more general in the well-stratified ocean-interior, while convection is known to occur near the sea-surface, e.g. via nighttime cooling. Far below the surface, the Western-Mediterranean Sea is very weakly stratified and offers opportunity to observationally study deep-sea convection. An opportunistic small set of high-resolution temperature sensors demonstrates not only classic internal-wave-induced turbulence, but also convection attributed to geothermal heating and spectral properties that relate to various chaos-theory models such as $1/\sigma$ pink noise ($\sigma$ denoting frequency), mainly found lying at (0.01 m above) the seafloor, and $1/\sigma^2$ Brownian noise, mainly found on a moored line at about 100 m above the seafloor. Near-inertial temperature variations are observed to occur down to the seafloor thereby disturbing the local convective turbulence regime to shear-dominated one temporarily. The integral turbulence time-scale is generally smaller (with dominant higher frequency motions) at the seafloor than about 100 m above it.

**I. INTRODUCTION**

Although the ocean and deep-seas are generally very turbulent in terms of large bulk Reynolds numbers Re well exceeding Re > $10^4$ and more generally Re = $O(10^6)$, it is a challenge to study the dominant turbulence processes. As the ocean is mainly stably stratified in density, which hampers the size-evolution of fully developed three-dimensional isotropic turbulence, it is expected that in the deep-sea stratification is much weaker resulting in near-neutral conditions of (almost) homogeneous waters. Near-homogeneous weak stratification conditions are considered in which buoyancy frequency N ≈ f, the local inertial frequency of Earth rotation. Under such conditions turbulent overturns may be slow and large, and may govern more the convection-turbulence process than shear-turbulence process that dominates under well-stratified conditions.

Besides these two characteristic turbulent (mixing) processes, stratification supports internal waves which may transfer their energy to irreversible turbulence after becoming non-linear. A characteristic of internal wave occurrence is 'intermittency', which is mathematically at the edge of deterministic chaos (Schuster, 1984) and may reflect the non-linearity of internal waves. The relationship between intermittency and turbulence has not often been studied in deep-sea observations.

In this paper, rare deep-seafloor high-resolution temperature (T-)measurements are analyzed and compared with data from T-sensors some distance above the seafloor in the deep Western-Mediterranean Sea. The goal is to study the variations with time of dominant turbulence types, shear- or convection-induced, in the deep-sea. Also, the importance of different sources like near-inertial waves is investigated that may be generated locally, after geostrophic adjustment of frontal collapse, or remotely, e.g. after the passage of atmospheric disturbance. The T-sensors are opportunistically retrieved during a deep-sea salvage-operation after 18 days of sampling. The locally extremely weak stratified conditions vary with time, which offers the opportunity to study various turbulence processes. As will be demonstrated, frictional flow processes induce only weak turbulence, compared with that from internal waves, eddies, and convection, especially also at the seafloor. This is because the weakly stratified waters set the condition for (the observation of) geothermal heating from below. Geothermal heating is defined as the general heat flux due from the Earth's mantle through its crust, outside underwater volcanic thermal vents. Generating predominantly convection-turbulence, geothermal heating is an effective vertical mixing process. In the Western-Mediterranean, it contributes about 0.1 W m$^{-2}$ of heat flux to the overlying deep-sea water (Pasquale et al., 1996).

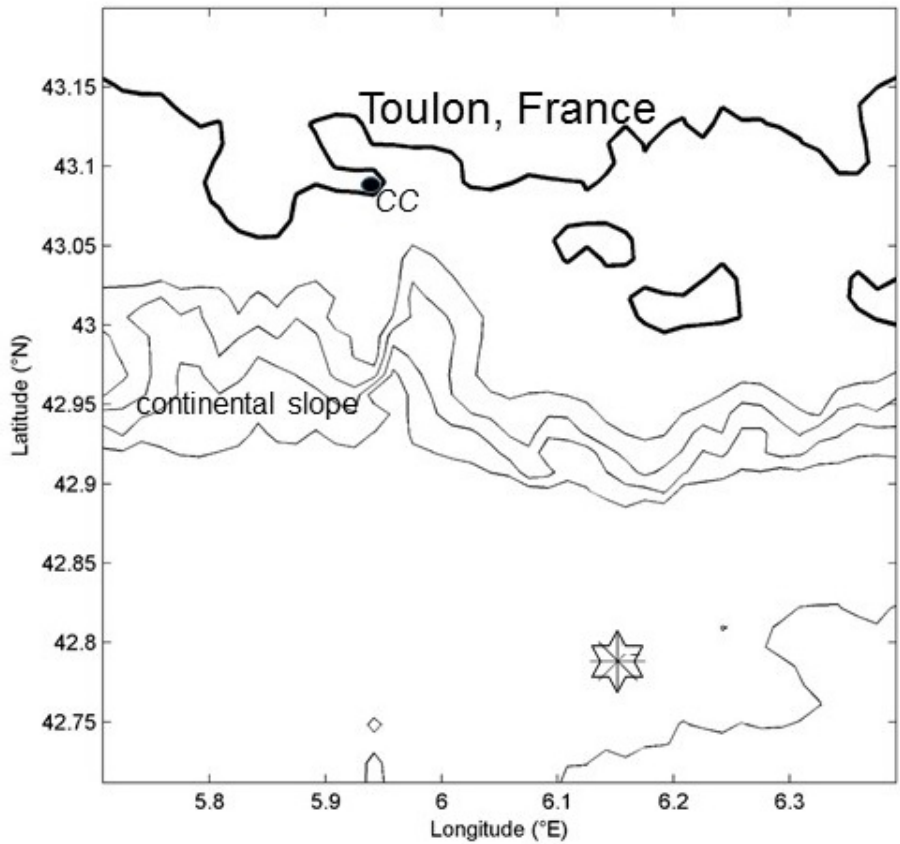

**Figure 1**. Mooring location (star) at the northern side of the Provençal Basin of the Western-Mediterranean Sea, about 40 km south of Toulon, France. Isobaths are drawn every 500 m. CC indicates Cap Cepet, the weather station at the entrance of the Bay of Toulon.

**II. MATERIALS AND METHODS**

**A. Site and data**

For a study on 3D-aspects of turbulent overturn development in the deep Mediterranean Sea, a 70-m diameter, 125-m tall, 45 vertical-lines mooring array has been deployed on 09 October 2020 at 42° 49.50´N, 06° 11.78´E, H = 2458 m water depth, on a flat sedimentary plane about 10 km south of the foot of the steep continental slope and about 40 km south of Toulon, France (Fig. 1). The array holds a total of 2925 renewed high-resolution 'NIOZ4n' T-sensors that sample half a million liters of seawater at a rate of 0.5 Hz for a period of about 3 years. Details on its construction and deployment can be found in van Haren et al. (2021). In this paper, 18-day records are analyzed from four T-sensors that were partially opportunistically retrieved during an underwater operation using a Remotely Operated Vehicle (ROV). This operation occurred on 19 and 20 November, about 5 weeks after deployment of the ring, and about 2.5 weeks after the automatic start of the T-sensor sampling on 01 November 06 UTC. The ring is intended to stay underwater for 3 years.

As the mooring array was deployed in free fall with a drag-parachute for stabilization, only a post-deployment ROV-operation could reveal the result of proper landing and the self-unrolling of the individual mooring lines after resolution of chemical releases 5-7 days underwater. The ROV-operation became acute, because two (after 10 days underwater: one) of the six drag-parachute lines became stuck and could not be released acoustically from the research vessel. The stuck drag-parachute was eventually released by the ROV, and at least one vertical mooring line was partially damaged by the stuck drag-parachute. This resulted in some scraping-off of T-sensors that were originally taped to the lines: From the nearest line at 2.5 m horizontally away, the stuck drag-parachute caused 9 sensors to fall-off and 5 to move along the line, all in the upper one-third of the 125-m long line originally holding 63 T-sensors. All fallen-off sensors were lying still horizontally so that their (cage-protected) sensor-tip was 0.5-1 cm from the sediment.

Unfortunately, the ROV could not operate within the large-ring perimeter of mooring lines that were 9.5 m apart horizontally, because of its umbilical cord to the research vessel 2500 m above. On the off-chance, a few T-sensors fell outside the large ring and could be retrieved from the seafloor by the ROV. One sensor was at the seafloor at least 2 m from the 0.6-m diameter steel large-ring pipe from the start of measurements, and presumably fallen-off just after the chemical release of its vertical line. Another one, which was originally at 97 m above the seafloor, fell off the vertical mooring line after ROV-release of the drag-parachute.

All 6 drag-parachute were equipped with old 'NIOZ4o' T-sensors. Four of these sensors could be retrieved and of which two carried undamaged records: One from the stuck acoustic release at 1 m above the seafloor and the other floating at about 140 m above the seafloor. These records will be discussed below as reference for the two described above.

NIOZ4 are self-contained T-sensors with a precision better than $5\times10^{-4}$°C and a noise level of less than $1\times10^{-4}$°C (van Haren, 2018). Every 4 hours when on a mooring line, all sensors are synchronized via induction to a single standard clock, so that times are less than 0.02 s off. This is not the case here for the T-sensors that were taped on the drag-parachute releases, and they had to be synchronized manually. T-sensors are calibrated to within noise level in a laboratory thermo-bath, but typical electronic drift is about $1\times10^{-3}$°C/mo after aging. When many T-sensors are densely mounted on a vertical mooring line, the drift is corrected by fitting to smooth stably stratified vertical profiles. Because this drift-correction is not possible for the present data from four sensors, these records are aligned to their minimum values. As the focus is on spectral analysis however, absolute values are not relevant. Drift of the two T-sensors from the vertical mooring line could be corrected properly because they were together in the deep nearly homogeneous waters in the same basket of the ROV after retrieval.

With respect to the NIOZ4o mounted on the drag-parachute lines, the NIOZ4n mounted on the large mooring array are expected to have lower noise level, less electronic drift and less spiking due to data storage failure. After pressure and drift correction (for these data: manual-slope correction and noise averaging of the first 1000 data-points), Conservative Temperature (IOC, SCOR, IAPSO, 2010), henceforth 'temperature' for short.

To establish the local vertical temperature, density and buoyancy frequency profiles, one shipborne SeaBird-911 Conductivity-Temperature-Depth (CTD) profile was obtained to 0.5 m from the seafloor and 1 km West from the mooring site on the day of deployment. Three-hourly meteorological data were available from Cap Cepet, the nearest station of Meteo France (Fig. 1).

**B. Spectral slopes**

An important tool in studying internal waves and turbulence using Eulerian scalar temperature observations is their representation in spectral form. Below, several spectral forms are reviewed for ocean observations.

As the Fourier transform underlying the transformation of a time series to a spectrum of a frequency series is a decomposition of the signal in sines and cosines it highlights a deterministic signal with a particular fixed amplitude and phase. Because the decomposition is only statistically significant for non-existing infinitely long time series, relevant statistics for realistic finite time series are obtained via smoothing the spectrum, either by averaging the variance contents of neighboring frequency bands or by averaging spectra of a number of (best half-overlapping) shorter time series (Emery and Thompson, 1998). While this exercise broadens the spectral peak of deterministic signal by distributing its energy over a larger fundamental bandwidth it is fundamentally irrelevant, as a deterministic signal has known amplitude and phase by definition and does not need quasi-random statistics to establish its significance (van Haren 2016). Spectral statistics are relevant for the type of signals they represent: quasi-random signals, or white noise. The question is, how spectral statistics should be interpreted for intermediate signals, such as internal tidal wave signals that have a deterministic source but become modified by non-deterministic variations in background stratification and wave-wave interactions. Of concern are the averaging effects on relatively broadband signals of the transition from internal waves to turbulence which are characterized by their spectral slopes (in a log-log plot). Spectral averaging shows the dominant signal or slope, but may affect the interpretation.

Various physical and mathematical models of frequency spectra are distinguished of a scalar like temperature in a stratified turbulent environment. For relatively strongly stratified waters of which $N > 10f$, the band of propagating internal gravity waves $f \leq \sigma \leq N$ scales like $\sigma^{-2}$, or a slope of -2 in a log-log plot, at frequencies well away from its boundaries $f \ll \sigma \ll N$ (Garrett and Munk, 1972) outside near-inertial and near-buoyancy cusps and humps, respectively (Munk, 1980). This canonical internal wave model does not account for areas with strong internal tidal harmonics as it was fit to data from the Mediterranean and Western Atlantic Ocean, where internal tides are relatively weak. A -2-slope also reflects fine-structure, or the advection of infinitely small step-functions passing a sensor (Phillips, 1971; Reid, 1971). The convolution of a finite time series of a Heaviside step-function constructs a -2-slope. A -2-slope also reflects Brownian motion (Schuster, 1984).

It may be shown that finitely small steps advected by lower frequency internal waves passed a sensor yield steeper slopes up to -3 (Gostiaux and van Haren, 2012). This slope manifests itself at frequencies $\sigma > 1.6N$, as the advection is by all internal waves, in an environment of the open ocean far away from boundaries that is weakly turbulent. In such environment, the background internal wave band slopes like -1, which may represent an unsaturated internal wave field. Extensions of it to super-buoyancy frequencies have been observed deep in the Mariana Trench (van Haren, 2020). Mathematically, a -1-slope also represents an intermittent signal with long regular phases and short irregular bursts (Schuster, 1984). Such regime is weakly chaotic, or at the edge of chaos and is e.g. modelled by non-linear dynamical systems having (the transverse) Lyapunov exponent equal to zero (Ruseckas and Kaulakys, 2013). In historic observations from the stable nocturnal atmospheric boundary layer (Caughey, 1977), a -1-slope is visible, albeit unnamed, at lower frequencies adjacent to a -5/3 of turbulence inertial subrange. Such a -1-slope is also visible in observations made under unstable atmospheric boundary-layer conditions (Williams and Paulson, 1977).

In a stronger turbulent environment the Kolmogorov model describes a forward cascade of energy from its source at low frequencies to dissipation at high frequencies with an equilibrium inertial subrange in between (Kolmogorov, 1941; Ozmidov, 1965). For a passive scalar representing shear-induced turbulence the inertial subrange slopes like -5/3 (Obukhov, 1949; Corrsin, 1951; Ozmidov, 1965; Tennekes and Lumley, 1972). This reflects a passive scalar (Warhaft, 2000). In a stratified turbulent environment the transition from source, e.g. internal waves, to dissipation (into heat) may also incorporate a buoyancy subrange, in which turbulence is anisotropic and which is bounded by the Ozmidov frequency $\sigma_O = U/L_O$ of largest possible isotropic length scale $L_O$ in a stratified environment, and/or a convective subrange which for an active scalar slopes like -7/5. This is known as the Bolgiano-Obukhov scaling (Bolgiano, 1959; Pawar and Arakeri, 2016) for an active-scalar buoyancy-driven convection-turbulence, which has been observed in laboratory and alpine Lake Garda at super-buoyancy frequencies $N < \sigma < \sigma_O$ (van Haren and Dijkstra, 2021).

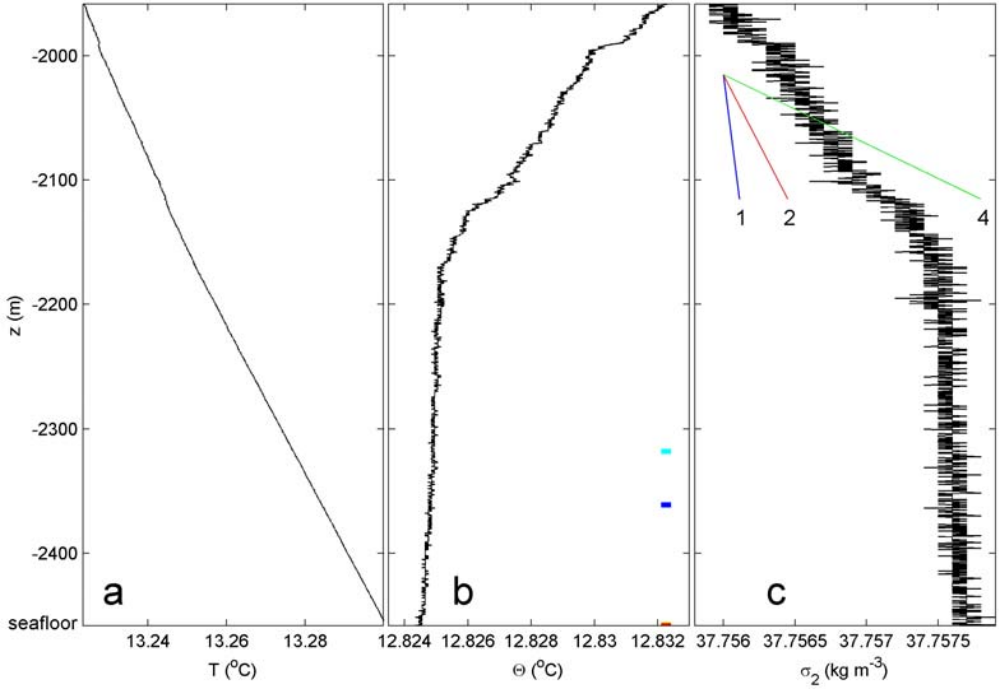

**Figure 2**. Shipborne CTD-profile obtained during mooring deployment to within 0.5 m vertically from the seafloor. Data are plotted between the seafloor at z = -2458 m and 500 m above it. (a) Temperature. (b) Conservative Temperature. The coloured ticks indicate the vertical position of the four T-sensors (the yellow one at 1 m above the seafloor is barely visible under the red one at the seafloor). (c) Density anomaly referenced to $2\times10^7$ N m$^{-2}$. The three sloping lines indicate the buoyancy over local inertial frequency ratio N/f and are a measure for the vertical density stratification

## III. RESULTS

### A. Vertical profile

The shipborne CTD observations demonstrate weakly stratified but not completely homogeneous waters in the lower 500 m above the seafloor (Fig. 2). In the lower 250 m above the seafloor of this single profile, vertical temperature variations are dominated by pressure effects and basically represent the adiabatic lapse rate (Fig. 2a). Conservative Temperature and density anomaly relate consistently with monotonic decrease and increase with depth, respectively (Fig. 2b, c). Over a vertical range of 100 m, density stratification varies so that N ≈ 1f near the seafloor, N ≈ 2f around z = -2050 m. Over 1-10 m small vertical ranges, N ≈ 4f and larger can be found, like around z = -2110 and -1980 m (Fig. 2c). Such thin stratified layers are occasionally also found at greater depths if the better-resolved temperature profile is investigated (Fig. 2b). Although the moored T-sensors were located between the seafloor and z = -2318 m, stratification may vary with depth and time.

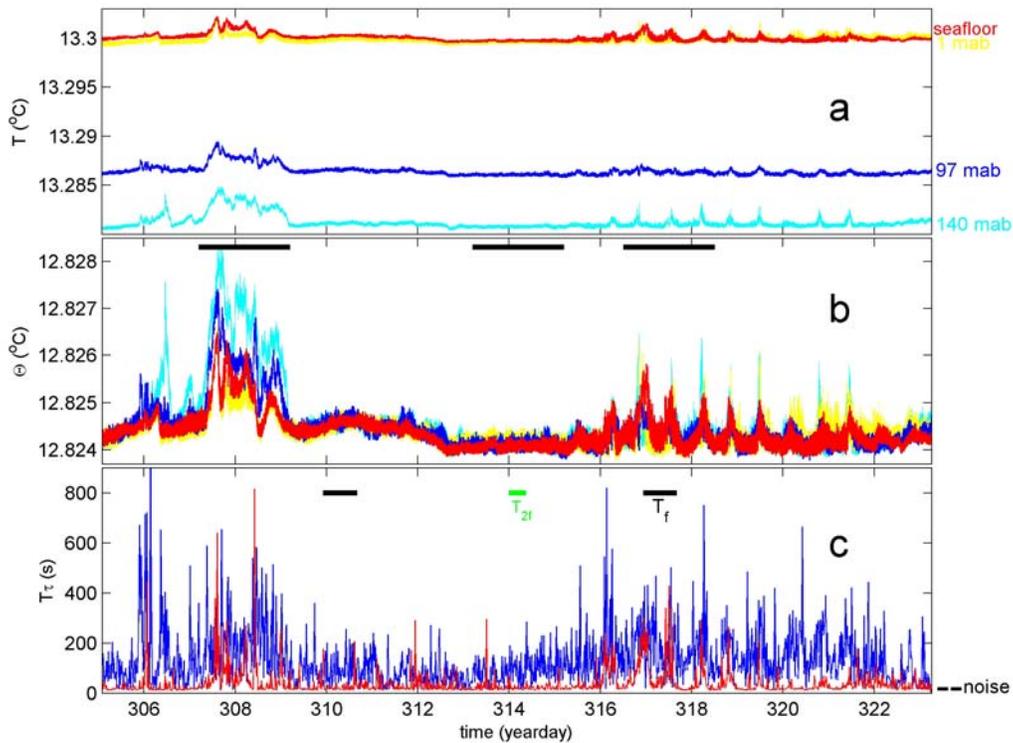

**Figure 3**. Time series overview of temperature 'T-'sensor data. Renewed-sensor NIOZ4n-data are in red (sensor at the seafloor) and blue (sensor on mooring line, about 97 m above bottom (mab for short). Older-sensor NIOZ4o-data are in yellow (sensor attached to stuck release, at 1 mab) and in light-blue (sensor attached to released parachute line, about 140 mab). (a) Temperature, calibrated but not drift-corrected. (b) Conservative Temperature, with NIOZ4n-data drift-corrected (see text) and NIOZ4o-data corrected with respect to the minimum value in the record. The 2-day horizontal black bars indicate the times of short-period spectra in Fig. 5. (c) Integral turbulence timescale (see text) computed from 2000-s sections of NIOZn-data only. The horizontal black bars indicate the inertial period, the green bar the semi-inertial period.

**B. Time series**

The two-and-a-half week time series of moored T-sensor data demonstrate considerable variability (Fig. 3). The original calibrated but otherwise uncorrected temperature records confirm the positions of the sensors in weakly stratified waters in which pressure dominates temperature variations: The average highest temperatures are found at the two T-sensors at, and 1-m from, the seafloor, whereas the average coolest temperature is found at the highest T-sensor, at about 140 m above the seafloor (Fig. 3a). The mean temperature differences are close to the local adiabatic lapse rate, but not exactly so due to the lack of drift-correction. Drift is visible in the two records from near the seafloor, with a more constant record over the long-term for the NIOZ4n T-sensor at the seafloor and a slightly sloping record at about 0.0006°C in 18 days for the NIOZ4o T-sensor at 1 m above the seafloor. The same discrepancy is found for the two T-sensors higher up, which evidences less drift in NIOZ4n compared to NIOZ4o T-sensors.

Corrected records demonstrate a temperature-variations range of 0.004 °C (Fig. 3b). In the 18-day time series, largest temporal and vertical variability are found between days 306 and 309, a relatively quiescent period with small variability is observed between days 309 and 315, and modest variability with dominant inertial periodicity between days 315 and 322 (Fig. 3b).

Between days 306 and 309, stratification is evident, with largest temporal variability at the uppermost T-sensor. Over the range of observation, $N \approx 2f$ and peaks up to $4f$, judging from these records that coarsely resolve the vertical. It is hypothesized that the warmer, more stratified waters come from above or are advected horizontally, and reflect the CTD-profile of Fig. 2b with its stratification above $z > -2170$ m going deeper. These waters are related with the passage of (sub-)mesoscale eddies governed by the (dynamically unstable) boundary flow along the continental slope (e.g., Crepon et al., 1982). In mid-autumnal November, the boundary flow and eddies are not related with deep dense-water formation, which is found to occur in the area but in late-winter and early-spring only. Evidently, such eddies reach all the way down to the deep seafloor. Models have demonstrated that cyclonic two-dimensional eddies or vortices are stabilized by three-dimensional turbulence, whilst anti-cyclonic vortices are disrupted (Bartello et al., 1994).

During the other two periods, stratification seems considerably less, being close to neutral, and episodically even (apparently) unstable, especially during the inertial-dominated period between days 315-322. Curiously, around day 314 a semi-inertial ($\sigma = 2f$) periodicity is discernible. The manifest inertial periodicity between days 315-322 in the temperature record from the sensors at and near the seafloor demonstrates the effects of possibly atmosphere-induced inertial motions reaching the (flat) seafloor or by more local generation, e.g. of collapsing fronts. Motions at near-inertial frequencies follow from geostrophic adjustment of a passing disturbance (Gill, 1982). Such a disturbance may cause immediate barotropic (over the entire water depth) near-inertial response at depths of around 1000 m (Shay and Elsberry, 1987) and even deeper than 4000 m (Morozov and Velarde, 2008) due to sea-level variations. Secondary response is by downward propagating density-gradient-driven baroclinic near-inertial waves is much slower, typically 10-15 days, with quasi-horizontal motions in stratified waters (Morozov and Velarde, 2008), which, in an unusual way can attain a non-negligible vertical component of dominant gyroscopic waves under neutral conditions (LeBlond and Mysak, 1978; van Haren and Millot 2004). It is unclear why the inertial amplitude is smallest under neutral conditions at some distance (97 m) from the seafloor, which was confirmed by a similar record of a sensors fallen down from 85 m above the seafloor (not shown).

The small-scale variability of motions above instrumental (white) noise is investigated in terms of the integral (turbulence) timescale $T\tau$, which is defined from the normalized cross-correlation $r(\tau) = <T'(t)T'(t+\tau)>/<T'^2>$ (e.g., Kundu and Cohen, 2008) for temperature fluctuations in blocks of 1000 data-points (2000 s) as,

$$T\tau = \int r(\tau)d\tau, \quad r(\tau) > 0.5, \tag{1}$$

but only for the lower noise-level NIOZ4n T-sensors. T$\tau$ represents the time over which the process T′(t) is highly correlated to itself. When this time-scale is large, 'low-frequency' large-scale more-energy-containing turbulent overturns dominate. When small, correlation is rapidly lost and 'high-frequency' small-scale turbulent motions dominate.

The computed time series (Fig. 3c) demonstrate generally increased T$\tau$-values when large-scale temperature variability (cf. Fig. 3b) is larger, but the largest T$\tau$-values do not coincide precisely with large-scale temperature variations. In general, T$\tau$-values are increased for the T-sensor on the mooring line 97 m above the seafloor and only occasionally reach noise levels (T$\tau \approx$ 20 s). The red T$\tau$-record from the T-sensor at the seafloor more often reaches noise level, which reflects a lack of dominant energy-containing turbulent overturns.

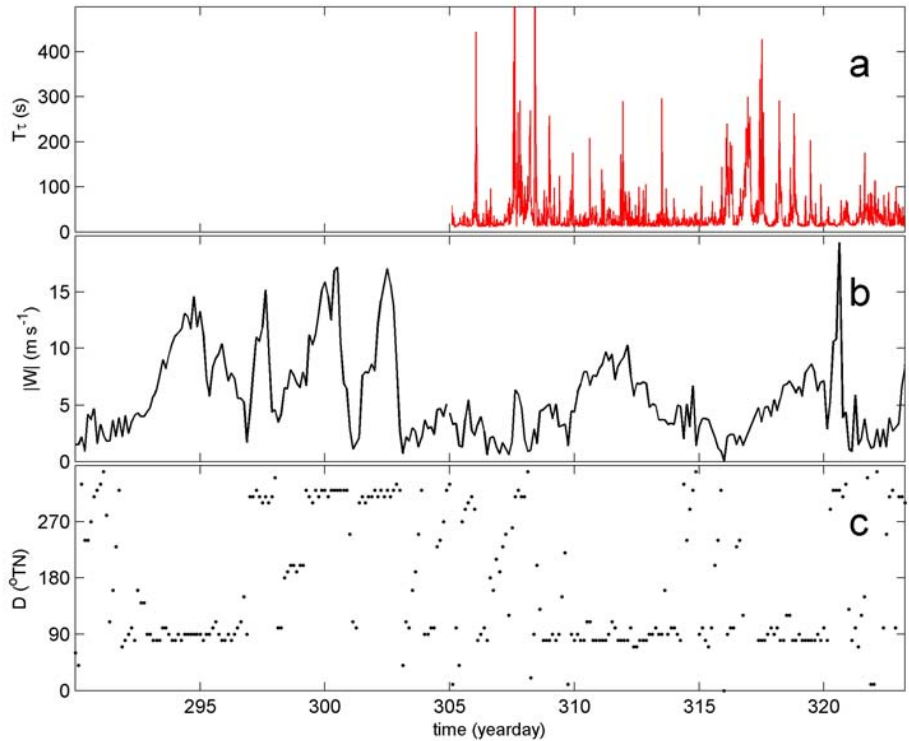

**Figure 4**. Potential correspondence between local wind, measured at Cap Cepet about 35 km Northwest of the mooring site (Fig. 1), and deep turbulence. (a) Integral turbulence timescale from Fig. 3c (sensor at the seafloor). Note the smaller y-axis range compared to Fig. 3c. (b) Wind speed. (c) Wind direction.

In both T$\tau$-time series, in particular also at the seafloor, a distinct inertial periodicity is seen, with larger T$\tau$ when temperature is highest in the inertial period. In the T$\tau$-record from the T-sensor at the seafloor, peaks at the inertial periodicity also occur during the first 10 days of the record, when inertial periodicity was less evident in the temperature records in Fig. 3b. This continued T$\tau$ inertial periodicity further demonstrates the effects of possibly atmosphere-induced inertial motions reaching all the way to the deep seafloor, in such a way that the originally horizontal motions near the surface where stratification is relatively large must have attained a non-negligible vertical component to show in the present temperature time series under near-neutral conditions. The T$\tau$ of Fig. 3c show that deep-sea inertial motions affect turbulence (periodicity), either directly at the inertial period or, as e.g. around day 314, at its first harmonic semi-inertial period.

From Fig. 4, the potential correspondence is investigated between local winds at the sea surface and turbulence at the seafloor induced by near-inertial motions. Winds are plotted up to 15 days prior to the start of the mooring measurements. No direct (barotropic) correspondence is observed between generally easterly winds and the group of inertial turbulence periodicity between days 315 and 320. If these turbulence variations are initiated by the impulse of north-westerly winds between days 297-303, it would have taken the (baroclinic) near-inertial waves about 15 days to reach the seafloor, which corresponds with observations in the Pacific Ocean by Morozov and Velarde (2008).

**C. Frequency spectra**

The 18-day overall mean temperature variance spectra scaled with turbulence inertial subrange $\sigma^{-5/3}$ show a peak about 10% higher than f and a minor one at 2f (except at 97 m above the seafloor which shows a spectral dip). Indeed, non-significantly higher values are found at the seafloor (Fig. 5, upper panel). Outside the inertio-gravity (combined internal gravity and gyroscopic) wave (IGW) band [$\sigma_{min}$, $\sigma_{max}$] (LeBlond and Mysak, 1978) for $\sigma > \sigma_{max}$, the different (turbulence) spectra diverge.

Although not well resolved, the IGW-peaks seem to fall-off in variance following slope like -1/3 (which is -2 in an unscaled log-log plot), which continues to about 5 cpd ($\approx$ 4f, the small-scale buoyancy frequency maximum). Henceforth, cpd is used for 'cycles per day'. Between $\sigma_{max}$ (or 4f) and 10 cpd, all spectra slope like +2/3 (-1 in unscaled log-log plot). At 10 cpd, a monotonic increase in temperature variance exists with distance from the seafloor.

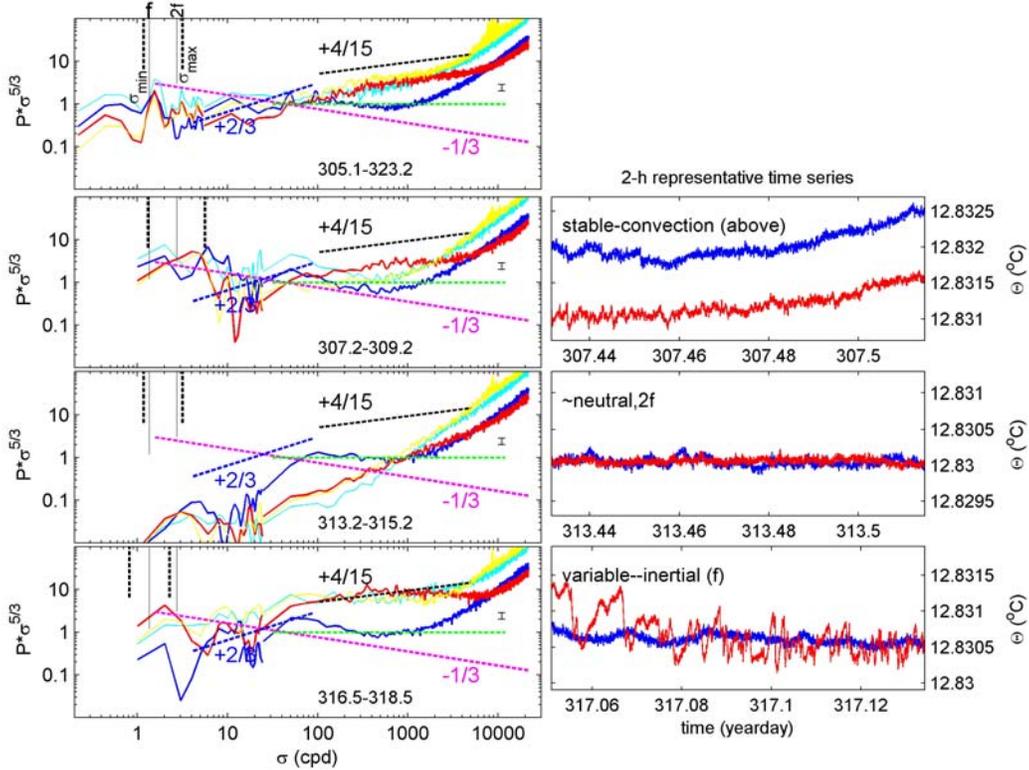

**Figure 5**. Internal wave-turbulence information spectral temperature information with two-hour representative time series for Fig. 3b-data using the same colors. The spectra are moderately smoothed, about 20 degrees of freedom (dof) near the inertio-gravity wave (IGW) range [$\sigma_{min}$<f $\sigma_{max}$>N] and heavily smoothed (about 800 dof) for the turbulence range of about [10 1000] cpd with error bar indicated. Upper panel shows the entire 18-day mean spectra with IGW-range (heavy dashed lines) for N = 2f. Besides varying IGW-range, f and 2f frequencies are indicated (thin solid lines) in each spectral panel. T-variance is scaled with the inertial-subrange spectral power slope $\sigma^{-5/3}$. On the log-log plot spectral slopes are represented by straight lines, with zero-slope for shear-turbulence (green-dashed; -5/3 in unscaled plot), +4/15 for convective turbulence (black-dashed; -7/5 in an unscaled plot), +2/3 for intermittency noise (blue-dashed; -1 in an unscaled plot), -1/3 for Brownian (red) noise or internal waves (purple-dashed; -2 in an unscaled plot). Three distinct two-day periods of time indicated by yearday-numbers below represent: Stable convection period from above with IGW for N = 4f, near-neutral period with weak semi-inertial variations and IGW for N = 2f, and variable weakly stratified period with dominant inertial variations and IGW for N = 1f.

Between well resolved 10 < $\sigma$ < 100 cpd, the two upper sensors attain more or less zero-slope (-5/3 in an unscaled plot), while the lower two continue to slope like -1. This is well outside the IWB, but may reflect intermittent signals. For well resolved $\sigma$ > 100 cpd up to rolling-off to their respective noise levels of +5/3 slope (0 in an unscaled plot), the uppermost and the lower two T-sensors attain a slope of -4/15 (-7/5 in unscaled plot). The mooring-line spectrum from 97 m above the seafloor slightly slopes downward, close to -1/3 before resuming zero-slope. At about 700 cpd, its variance is more than half an order of magnitude lower than that of the other records.

Roll-off to instrumental noise levels start at about 800 cpd for upper parachute-line and mooring-line sensors, 1500 cpd for the sensor at the release at 1 m above the seafloor and about 6000 cpd for the sensor at the seafloor. The two NIOZ4n T-sensors have lower noise level, as at the Nyquist frequency their variance is about 3 times lower than that of the two NIOZ4o T-sensors. This is a factor of 1.7 lower noise level in temperature (one standard deviation). The different T-sensors of each (custom-)make also have slightly different noise levels, by chance.

This divergence in turbulence-spectral slopes reflects a dichotomy of dominant shear-induced turbulence at 97 m above the seafloor and dominant convective turbulence near the seafloor and 140 m above. The record of the T-sensor at the seafloor shows +4/15-slope over a frequency-range of nearly two orders of magnitude, without reaching into the inertial subrange before or after. However, there are shorter episodes when this record also shows evidence of shear-induced turbulence, which thus occasionally reaches the seafloor, as will be demonstrated below.

The 18-day record is split in three two-day periods that represent the three different episodes of different temperature variability, as indicated in Fig. 3b. Two-day periods naturally do not resolve the IGW-range well, but the turbulence-range of $\sigma > 10$ cpd is well resolved with heavy smoothing. During the largest temperature and stratification variations around day 308, which are reflected in largest variance in and just above IGW, the two NIOZ4n-records most reflect the 18-day spectrum for the turbulence-part of $\sigma > 10$ cpd. For this episode, both sensors at some distance above the seafloor bend into the inertial subrange from a -1/3-slope, with about half an order of magnitude more variance in the uppermost sensor-record. The record at a m above the seafloor is also differs strongly from the 18-day mean spectrum, being similar to the one from the seafloor-sensor in the IGW, but roughly following zero-slope between 10 and 600 cpd before rolling off to noise when equal in variance with the uppermost sensor.

During the period of weakest variance around day 314, IGW-variance is relatively low, and three out of four spectra slope +2/3 up to the frequency where they reach their instrumental noise level. No inertial subrange slope is found in these spectra. In the IGW, they show a bulge around 2f-$\sigma_{max}$. The spectral exception is the NIOZ4n T-sensor at 97 m above the seafloor, which show a larger super-IGW bulge and strictly follows zero-slope between about 60 and 1000 cpd, before rolling off to noise.

During the inertial-frequency dominated period around day 317, all spectra show a moderate peak just higher than f and a less clear one just higher than 2f (if a 5-day period spectrum is computed for sufficient resolution). Most IGW-variance for the sensor at the seafloor and least IGW-variance for the sensor at 97 m above are also visible in the plotted 2-day spectrum. While all four records show +2/3-slope for [6, 40] cpd, only the record from 97 m above the seafloor resumes with -1/3-slope for $\sigma > 40$ cpd. A zero-slope is visible in all spectra between about 300 cpd and their rolling-off to noise. This slope starts at about 40 cpd for the two NIOZo-sensors, and is preceded by a -4/15 slope for the seafloor-sensor. Their variance is close to identical, being the highest turbulence-level in the 18-day record, and up to one order of magnitude larger than found in the spectrum from 97 m above the seafloor that shows an almost constant level of zero-slope between 200 and 800 cpd for all periods.

This large turbulence-variance discrepancy shows in the 2-h, 0.002-°C range plots: In general low-frequency motions are slightly dominant at 97 m above the seafloor with unknown 0.01-day periodicity, whereas high-frequency turbulent motions dominate at the seafloor. It is noted that none of the short-range time series show instrumental noise levels, which has one standard deviation of 0.00003°C for the T-sensor at the seafloor.

**IV. DISCUSSION AND CONCLUSIONS**

The comparison between temperature records from high-resolution sensors at the deep seafloor and on a mooring line about 100 m above demonstrate considerable variability in turbulence characteristics. The 18-day time series show three distinguished periods with different potential turbulence sources under weakly stratified conditions: a stable stratification advected from above, and two near-neutral conditions with dominant semi-inertial and inertial temperature variability.

As tidal motions are small the Western-Mediterranean Sea mainly, the dominant IGW are near-inertial motions which are a result of disturbance-adjustment under Earth rotation. While well-known disturbances are passages of atmospheric fronts and depressions, inertial motions may be generated locally following passages of (collapsing) underwater fronts or horizontal density differences. Under near-homogenous weakly stratified conditions, near-inertial waves are the only IGW that can propagate freely from stratified to neutral waters and vice versa (e.g., LeBlond and Mysak, 1978). However, linear waves do not induce turbulence, for which non-linear interactions are needed to the point of wave-breaking. In near-homogeneous waters, (small-scale) eddies and possibly also wave-propagation may transfer to slantwise convection in the direction of the Earth rotation (e.g., Straneo et al., 2002; Sheremet, 2004). This will have effect on the character of deep-sea turbulence.

The observed difference in internal wave-turbulence character between the unique T-sensor at the seafloor and higher-up is summarized as follows:

-Inertial temperature variations are generally larger at the seafloor than above.

-At super-IGW frequencies, the spectra dip (but do not show a gap), generally deepest at or near the seafloor, before resuming with an intermittency (-1-slope) character. The dip and the -1-slope are independent of the IGW-source, either being inertial motions or broad-band waves advected from above.

-Intermittency covers the entire turbulence range when IGW-forcing is weak, except at about 100 m from the seafloor.

-At 100 m above the seafloor an inertial subrange is always observed, independent from the turbulence source. This subrange is also observed at or near the seafloor under inertial-motions forcing.

-At the seafloor, inertial-subrange absence is either replaced by intermittency or, more dominantly, by convective subrange. Convection-turbulence is likely induced by general geothermal heating and overrules any frictionally-induced shear-turbulence.

-Turbulence coherence length-scales are generally shortest at the seafloor, except during the warming phase of an inertial period.

Future studies may focus on the extent of convection-turbulence from the seafloor, which seems to be mainly dominant at or close to the seafloor and not at 100 m above. Future studies may also focus on the apparent spectral overshoot leading to -2-slope before inertial subrange at 100 m above the seafloor. Further investigations are also welcomed on potential correspondence between surface winds and deep seafloor turbulence at near-inertial periodicities.

ACKNOWLEDGMENTS

This research was supported in part by NWO, the Netherlands Organization for the advancement of science. I thank NIOZ-MRF and the captain and crew of the R/V Pelagia for their very helpful assistance during construction and deployment.

AUTHOR DECLARATIONS

CONFLICT OF INTEREST

The author has no conflicts to disclose.

DATA AVAILABILITY

The data that support the findings of this study are available from the author upon reasonable request. Meteorological data are freely available from Meteo France at https://donneespubliques.meteofrance.fr.

**REFERENCES**

Bartello, P., Métais, O. and Lesieur, M., "Coherent structures in rotating three-dimensional turbulence," J. Fluid Mech. 273, 1–29 (1994).

Bolgiano, R., "Turbulent spectra in a stably stratified atmosphere," J. Geophys. Res., 64, 2226–2229 (1959).

Caughey, S. J., "Boundary-layer turbulence spectra in stable conditions," Boundary-layer Meteorol. 11, 3–14 (1977).

Corrsin, S., "On the spectrum of isotropic temperature fluctuations in an isotropic turbulence," J. Appl. Phys. 22, 469–473 (1951).

Crepon, M., Wald, L., and Monget, J. M., "Low-frequency waves in the Ligurian Sea during December 1977," J. Geophys. Res. 87, 595–600 (1982).

Emery, W. J. and Thomson, R. E., *Data Analysis Methods in Physical Oceanography* (Pergamon, Amsterdam, 1998), p. 634.

Gill, A. E., *Atmosphere-Ocean Dynamics*. (Academic Press, San Diego CA, USA, 1982), p. 680.

Gostiaux, L., and van Haren, H., "Fine-structure contamination by internal waves in the Canary Basin," J. Geophys. Res. 117, C11003, doi:10.1029/2012JC008064 (2012).

IOC, SCOR, IAPSO, *The international thermodynamic equation of seawater – 2010: Calculation and use of thermodynamic properties*. Intergovernmental Oceanographic Commission, Manuals and Guides No. 56 (UNESCO, Paris, France, 2010), p. 196.

Kolmogorov, A. N., "The local structure of turbulence in incompressible viscous fluid for very large Reynolds numbers," Dokl. Akad. Nauk SSSR 30, 301–305 (1941).

Kundu, P. K., and Cohen, I. M., *Fluid Mechanics, 4$^{th}$ ed.* (Academic Press, London, 2008), p. 872.

LeBlond, P. and Mysak, L. A., *Waves in the Ocean* (Elsevier, Amsterdam, 1978), p. 602.

Morozov, E. G., Velarde, M. G., "Inertial oscillations as deep ocean response to hurricanes," J. Oceanogr. 64, 495-509 (2008).

Munk, W. H., "Internal wave spectra at the buoyant and inertial frequencies," J. Phys. Oceanogr. 10, 1718-1728 (1980).

Obukhov, A. M., "Structure of the temperature field in a turbulent flow," Izv. Akad. Nauk SSSR, Ser. Geogr. Geofiz. 13, 58–69 (1949).

Ozmidov, R. V., "About some pecularities of the energy spectrum of oceanic turbulence," Dokl. Akad. Nauk SSSR 161, 828–831 (1965).

Pasquale, V., Verdoya, M., and Chiozzi, P., "Heat flux and timing of the drifting stage in the Ligurian–Provençal basin (northwestern Mediterranean)," J. Geodyn. 21, 205–222 (1996)

Phillips, O. M., "On spectra measured in an undulating layered medium," J. Phys. Oceanogr. 1, 1–6 (1971).

Reid, R. O., "A special case of Phillips' general theory of sampling statistics for a layered medium," J. Phys. Oceanogr. 1, 61–62 (1971).

Ruseckas, J., and Kaulakys, B., "Intermittency in relation with 1/f noise and stochastic differential equations," Chaos 23, 023102 (2013)

Shay, L. K., Elsberry, R. L., "Near-inertial ocean current response to hurricane Frederic," J. Phys. Oceanogr. 17, 1249-1269 (1987).

Sheremet, V. A., "Laboratory experiments with tilted convective plumes on a centrifuge: a finite angle between the buoyancy force and the axis of rotation," J. Fluid Mech. 506, 217–244 (2004).

Straneo, F., Kawase, M., and Riser, S. C., "Idealized models of slantwise convection in a baroclinic flow," J. Phys. Oceanogr. 32, 558–572 (2002).

Tennekes, H. and Lumley, J. L., *A First Course in Turbulence* (MIT Press, Cambridge, 1972), p. 300.

van Haren, H., "Philosophy and application of high-resolution temperature sensors for stratified waters," Sensors 18, 3184 (2018).

van Haren, H., "Do deep-ocean kinetic energy spectra represent deterministic or stochastic signals?" J. Geophys. Res., 121, 240–251, doi:10.1002/2015JC011204 (2016).

van Haren, H., "Challenger Deep internal wave turbulence events," Deep-Sea Res. I, 165, 103400 (2020).

van Haren, H., Dijkstra, H. A., "Convection under internal waves in an alpine lake," Env. Fluid Mech. 21, 305–316 (2021).

van Haren, H., and Millot, C., "Rectilinear and circular inertial motions in the Western Mediterranean Sea," Deep-Sea Res. I 51, 1441–1455 (2004).

van Haren, H., Bakker, R., Witte, Y., Laan, M. and van Heerwaarden, J., "Half a cubic hectometer mooring array 3D-T of 3000 temperature sensors in the deep sea," J. Atmos. Oceanic Technol. 38, 1585–1597 (2021).

Warhaft, Z., "Passive scalars in turbulent flows," Annu. Rev. Fluid Mech. 32, 203–240 (2000).

Williams, R. M., and Paulson, C. A., "Microscale temperature and velocity spectra in the atmospheric boundary layer," J. Fluid Mech. 83, 547–567 (1977).